\documentclass[twocolumn,11pt,twoside]{article}

\usepackage{fancyhdr,color, graphics, epsfig, graphicx, subfigure}
\usepackage{MC2R}
\usepackage{amsmath}
\usepackage{amsfonts}
\usepackage{graphicx}
\usepackage{subfigure}
\usepackage{url}
\usepackage{multirow}

\newcommand{\PPAYD}{PriPAYD}

\begin{document}

\title{Report on the ``Secure Vehicular Communications: Results and
Challenges Ahead'' Workshop}

\author{
\begin{tabular}{p{2in} p{1in} p{2in}}
\multicolumn{1}{c}{\textbf{Panagiotis Papadimitratos}} & &
\multicolumn{1}{c}{\textbf{Jean-Pierre Hubaux}} \\
\multicolumn{1}{c}{\textit{panos.papadimitratos@epfl.ch}} & &
\multicolumn{1}{c}{\textit{jean-pierre.hubaux@epfl.ch}}
\end{tabular}\\
Laboratory for Computer Communications and Applications \\
Swiss Federal Institute of Technology - Lausanne (EPFL)\\
Switzerland \\ }

\maketitle
\thispagestyle{fancy}
\normalfont

The Workshop on \emph{Secure Vehicular Communications: Results and Challenges Ahead} took place in February 20-21, 2008, on the EPFL campus, Lausanne, Switzerland. The event brought together experts, from a variety of organizations, working on vehicular communication systems, security and privacy. The fourteen presentations offered an overview of the latest results and reflected the views of public authorities, academia, and industry. During the one and a half days of the workshop, the thirty-five attendees had the opportunity to have an in-depth discussion on future research and development directions for vehicular communication systems security and privacy.

The developments in the area of vehicular networks and communication systems, and the increasing attention from industry, academia and authorities, motivated us to organize this workshop. Vehicular communications (VC), including vehicle-to-infrastructure (V2I) and vehicle-to-vehicle (V2V) communication, with the latter leading to vehicular ad hoc networks (VANETs), lie at the core of a number of research initiatives. They aim to enhance transportation safety and efficiency, with applications that provide, for example, warnings about environmental hazards (e.g., ice on the pavement), traffic and road conditions (e.g., emergency braking, congestion, or construction sites), and local (e.g., tourist) information.

Nonetheless, the unique features of VC are a double-edged sword: the rich set of tools they offer make possible a formidable set of abuses and attacks. Consider any wireless-enabled device that runs a rogue version of the vehicular communication protocol stack and injects forged messages or meaningfully modifies messages transmitted by vehicle on-board communication units; or a vehicle that forges messages in order to masquerade an emergency vehicle and mislead other vehicles to slow down and yield. Furthermore, it is possible for the vehicles and their sensing, processing, and communication platforms to be compromised. Worse even, it is not difficult to consider a node could
'contaminate' large portions of the vehicular network with false information: for example, a single vehicle can transmit false environmental hazard warnings that can then be taken up by all vehicles in both traffic streams. From a different point of view, consider a large number of wireless access points deployed across an urban area, or along a highway (at rest areas, gas stations, etc). With such a wireless infrastructure receiving transmissions from passing by vehicles, anyone that obtained access to such data could easily infer private information about the drivers and the vehicle passengers: their locations, their routes, their communications and transactions.

These simple examples of abuse indicate that in all circumstances vehicular communications must be secured and the privacy of their users should be protected. It appears that the security of VC systems and the protection of their users' privacy are indispensable. Otherwise, these systems could make anti-social and criminal behavior easier than it is today without the VC
technology. If this were the case, the benefits of deploying VC systems would be in jeopardy.

It is our belief that security and privacy concerns for vehicular communication systems should, and hopefully will, be addressed before the deployment of VC systems. It is our hope that this venue provided a survey of the state-of-the-art solutions, cross-pollinated research and development  efforts in two continents, increased further awareness, and thus contributed towards the objective of trustworthy vehicular communication systems.

\subsection*{Workshop Summary}

The workshop was opened by the remarks of Panos Papadimitratos and Jean-Pierre Hubaux of EPFL. The first session, chaired by P. Papadimitratos, set the stage, providing an update on \textbf{recent developments on vehicular communications and applications}, as well as a framework for the \textbf{coordination of efforts to secure vehicular communication systems}.

\emph{Wai Chen} of Telcordia delivered the first presentation, covering numerous aspects on the activities within the VII initiative of the US Department of Transportation (DoT), insights from deployment experience in Japan, implementation and field tests within the VII initiative, as well as a perspective on the role of security. The second talk, by \emph{Tobias Gansen} of AUDI, presented the point of view of his organization on VC-enabled applications that appear plausible and likely to be deployed. He termed these applications as ``Dreams,'' in contrast to ``Nightmares'' that are use-cases that can lead to significant problems, notably due to security. The first session was completed by the presentation of \emph{Antonio Kung} of TRIALOG, which presented the scope and activities of the eSafety Security Working Group, which, co-chaired by A. Kung, has a role of providing recommendations to the European Commission for future research directions.

The late afternoon session, chaired by A. Kung, focused on \textbf{policy and standardization} issues related to vehicular communications and efforts to secure those systems. The first presentation, by \emph{Emilio Davilla-Gonzalez} of the European Commission, focused on policy and organizational challenges for VC security. The second talk was delivered by \emph{William Whyte} of NTRU, who also heads the security efforts of the IEEE 1609 working group on security. His presentation covered all the activities and latest developments within the VII initiative of the US DoT. The session closed by \emph{Benjamin Weyl} of BMW Research, who presented the activities and a roadmap of actions of the Car-to-Car Communication Consortium (C2C-CC) towards securing VC.

The second day opened with a session chaired by \emph{Bart Preneel} of K.U. Leuven. The two presentations in this session focused on \textbf{solutions and approaches to secure VC systems}. P. Papadimitratos covered the activities and results at EPFL and within the SeVeCom project; based on these results and on-going work, he discussed upcoming steps towards trustworthy VC systems. The
presentation of \emph{Christof Paar} of Ruhr University - Bochum concerned security applications in cars. In particular, C. Paar showed how cryptographic operations, such as signature verifications, can be done at high rates, as real-time support will be necessary in vehicular communications. He
also demonstrated how to break key-less car entry systems with a side-channel attack against a popular cipher and protocol.

The second morning session, chaired by \emph{Albert Held} of Daimler, offered presentations on \textbf{privacy enhancing technologies}. The first talk, by \emph{Bart Preneel} of K.U. Leuven, discussed concepts and products that enable pay-as-you-drive insurance, and then proposed and discussed the practicality of PriPAYD, a scheme that researchers at K.U. Leuven devised to protect the privacy of the insured driver. The second talk, by \emph{Thomas Heydt-Benjamin} of IBM Research, presented certain privacy and identity management mechanisms, notably those that are part of the systems developed in the context of the European project PRIME.

The rest of the day unfolded with a \textbf{demonstration session} and a single-talk session on \textbf{secure positioning}, both chaired by C. Paar. Demonstrations allowed the attendees to familiarize themselves with the PRIME prototype on location based services (presentation by T. Heydt-Benjamin), a network-traffic joint simulation tool TraNS (presentation by \emph{Maxim Raya}
of EPFL), and the DENSO hardware platform that is used by the VII initiative (presentation by \emph{Tim Leinmueller} of DENSO). After the demos, the talk on secure positioning was delivered by \emph{Neil Warfield} of GSA; he explained the concepts and latest developments towards deployment of the Galileo navigation system and the security services that Galileo is envisioned to offer.

The panel that followed had the topic \textbf{``Secure vehicular communications: What are the main research challenges left?''} and it was chaired by Jean-Pierre Hubaux. The panelists, W. Whyte, A. Held, and E. Davila-Gonzalez, proposed topics that need to be addressed in the future, towards a successful deployment of VC systems.

The closing session, chaired by \emph{Stefano Cocsenza} of CRF, presented developments on \textbf{in-car security} and \textbf{secure communication protocols}. The first talk, delivered by A. Held and prepared jointly with \emph{Thomas Eymann} of Bosch, covered requirements and approaches to develop solutions securing in-car communication systems. \emph{Andreas Festag} of NEC Labs presented certain mechanisms to secure geographical information assisted communication (Geocast) and enhance its privacy. The final presentation, delivered by \emph{Frank Kargl} of the University of Ulm, continued on secure vehicular communication, proposing a number of emerging related topics to
address.

In the rest of this paper, we provide a number of abstracts provided by the presenters themselves. The abstracts are presented in subsequent sections, in the order the corresponding presentations were made. We thank all the speakers for their contributions to this article and for their  presentations that are available on the workshop website.\footnote{\url{http://icapeople.epfl.ch/panos/SVCWCR/index.html}}

\section*{Some Recent Results on Vehicle Communications - Opportunities and Challenges}

\noindent \emph{Wai Chen, Telcordia Technologies}\\

Significant research efforts have been aimed at integrating communication and computing technologies into vehicles and roadway infrastructure. The objective is to improve preventive vehicle safety, reduce traffic congestion, and enable new applications such as diagnostics, mobile commerce, and entertainment. Industrial and governmental efforts are underway to accelerate the introduction of V2V / V2I communications functions including, e.g., the European C2C-CC and SeVeCom; the US VII and CAMP/VSC-2; the Japanese AHS / Smartway; and ASV, ISO/CALM, IEEE WAVE, and ETSI TC ITS.

Much of the recent research has been directed at seamless networking technology to effectively utilize heterogeneous communication media for vehicle users, and ad hoc networking technology for V2V and V2I communications. The combination of the requirements of emerging applications and characteristics of the roadway environments poses new challenges to the design of vehicular communication systems: to achieve high reliability, low latency, and data security in roadway environments. The mobility of vehicles can result in rapid network topology changes, node density fluctuations and constantly changing environment conditions. This could overwhelm the limited bandwidth of the radio links if the communication protocols are not well designed.

At the lower layers, there have been many efforts to design radio technologies that are tailored to communications in roadways. V2V channel modeling and effects on communications pose new challenges, given that vehicles,  both sending and receiving, can move at high speeds in roadway environments and the antennas are mounted at low vehicle heights. The vehicular applications require an efficient use of broadcast, multicast, and unicast in a heterogeneous network consisting of moving vehicles and stationary roadside units. A lot of efforts have been focused on designing MAC and network protocols among vehicles or roadside units that can support cooperative information downloading and emergency warnings distribution, among others. In terms of broad approaches, some efforts focus on one-hop broadcasting as the basic model; whereas others focus on using a vehicle group as a manageable unit for ad hoc communications to achieve controls over group size, message direction, and coordination (in transmission, routing, and multicast). Although feasible validation involving many vehicles remains a challenge, it is crucial to develop simulation capabilities for high-fidelity performance evaluations.

Initially, the densities will be low for equipped vehicles and roadside units that can be costly to deploy. Whereas preventive safety generally requires high-density of equipped vehicles to be effective, some experiments have shown that even low-density deployment levels can be beneficial to the reduction of traffic accidents.

Achieving security for applications and networking, and maintaining driver privacy in roadway environments are also crucial challenges, and as such have received much attention (e.g., the European SeVeCom and the US VII-C, among others).

\section*{Car-2-X Challenges - Dreams and Nightmares}

\noindent \emph{Tobias Gansen, Lars Wischhof, André Ebner, and Ingrid
Paulus, Audi Electronics Venture GmbH / AUDI AG}\\

The various projects in the area of Car-2-X communication have generated many different use-cases, which in turn have contributed to make the overall Car-2-X system complex and expensive. However, the rich variety of projects and initiatives such as the Car-2-Car Communication Consortium has successfully put this promising technology at center stage~\cite{c2c}. Customers, decision makers, 
as well as authorities show great interest in Car-2-X technology and expect nothing less than what the early visionaries of the field expected: the improvement of safety, mobility and comfort. It is now at us to make it real.

\subsection*{Car-2-X Use-Cases of Audi}

The Audi focus of Car-2-X use-cases is mainly on augmenting existing systems or sensors and on use-cases seeming to be relatively easy to introduce with immediate customer benefit. The usage of single hop broadcast in combination with a store-and-forward mechanism enables many delay tolerant use-cases such Decentralized Floating Car Data, Obstacle Warning or Vehicle Based Road Condition Warning. Simulations have shown that highly efficient systems are feasible with penetration rates as low as 2\% on highway scenarios~\cite{wis07}. Although these use-cases still require a minimum
penetration rate, Traffic Light and Signage Assistance will create customer benefit starting with the first traffic light equipped~\cite{men06}. More time critical safety use-cases like Pre-Crash Sensing and Preparation or Intersection Collision Warning are currently under investigation, but they require additional research not limited to technical examinations.

\subsection*{Use-Cases to Avoid}

Although many of the use-cases currently discussed are interesting on an academic basis, some of them are not attractive to OEMs. We as OEMs should have the strong desire to support the drivers of our cars, not to harass them. Therefore, all types of non-interactive safety inspections like Electronic License Plate should be avoided, from our point of view. This is also the case with the possibly cloaked introduction of a driver's log under the hood of the Car-2-X system. Such a system would not only add additional security requirements but may even discourage customers from choosing a Car-2-X enabled car. The definition of a flexible baseline security architecture \cite{deliverable2.1,sevecom2007} by the SeVeCom project and the detailed analysis of the different requirements~\cite{kro06} in combination with upcoming field operational tests can bring this technology one step closer to broad series introduction.

\section*{Vehicle Security in VII}

\noindent \emph{William Whyte, NTRU Cryptosystems, Inc}\\

Vehicle Infrastructure Integration (VII) is a US Department of Transportation (DoT)-sponsored initiative to enable vehicle-to-vehicle and vehicle-to-roadside communications in the 5.9 GHz band.\footnote{This work was supported in part by the US Department of Transportation.} The IEEE standards 1609.* and 802.11p standardize the communications stack. The first generation of IEEE standards was issued in 2006-7. And since mid-2006, the initiative has expanded from standardization work to a full-scale Proof of Concept (PoC) project, developing prototypes of both on-board equipment (OBEs) and roadside equipment (RSEs), as well as applications to run on the OBEs and RSEs and across the backhaul network. Field tests of the PoC systems have been underway since Q4 2007 and should conclude in Q2 2008.

IEEE 1609.2 is the standard for secure messaging in the VII setting.  The security work in VII PoC was based on 1609.2 but considerably widened the scope.

\subsection*{Types of Communication and Security Mechanisms in VII}

In addition to standard security requirements, in the VII system OBEs must have a guarantee of anonymity. This means that it should be difficult for an attacker, based on VII transmissions alone, to determine (a) that a specific transmission has come from a specific vehicle or (b) that two specific transmissions have come from the same vehicle (unlinkability). Typical communications scenarios include: 

\noindent \textbf{Traffic advisory} or \textbf{WSA multicast} from RSE - requires authentication. \emph{Supported by 1609.2 without modification.} 

\noindent \textbf{Safety-of-life multicast} from vehicle - requires authentication, anonymity. \emph{Required extension to 1609.2 to define anonymous mechanisms.}

\noindent \textbf{Tolling} - established by WAVE Service Advertisement (WSA), requiring RSE authentication); thereafter communications need confidentiality, authentication, and possibly non-repudiation. \emph{Addressed by the development of V-DTLS, a variant of Datagram TLS optimized for the VII setting to reduce round-trips and message size.}

\noindent \textbf{Lengthy communication with backhaul} - requires WSA authentication and an establishment of a secure session that can survive as the OBE moves between RSE communications zones. \emph{Addressed by the development of V-HIP, a variant of the IETF  Host Identity Protocol optimized to address anonymity requirements.}

\noindent \textbf{Communications with the CA} - \emph{Addressed by modifications and extensions to 1609.2.}

\subsection*{Additional Research}

Anonymity was supported by the so-called combinatorial method of creating N (= about 10,000) private key/certificate pairs and issuing n (= about 5) of those N pairs to each vehicle at random. Extensive simulation showed this to have good anonymity properties but to be fragile against large-scale attacks where the attacker compromises a large number of vehicles.

The certificate issuance mechanism separated the roles of issuing CA and authorizing CA; the two CAs must collude to compromise anonymity.

Extensive research was carried out on the mechanics for CRL distribution for OBE certificates.

Project participants also developed a cryptographic hardware accelerator capable of performing 250 ECDSA verifications a second.

\section*{eSafety Security Working Group}

\noindent \emph{Antonio Kung, TRIALOG} \\

The eSecurity Working Group is part of the eSafety forum, which brings together the European Commission, the industry, public authorities and other stakeholders to co-ordinate the advent of road safety applications. The WG was established in early 2007 to address the growing concern that the deployment of intelligent vehicle systems involves a large scale technology infrastructure
that is vulnerable to accidental or malicious misuse and therefore can jeopardize road safety. In particular:\begin{itemize}
\item Automotive industry manufacturers were concerned about vehicle intrusion problems, as the availability of permanent communication opens the door to a myriad of misuse cases that can threaten the integrity of vehicle electronics. 
\item Public authorities dealing with data protection issues were concerned that the availability and manipulation of  location-oriented data at such a high scale would create privacy problems. Without specific measures, the deployment of individual applications could lead to huge agreement overhead or, worse, to non-authorization recommendations.
\end{itemize}

The eSecurity Working Group can be viewed as a platform for European stakeholders to discuss these vulnerability aspects, with the following objectives: \begin{itemize}
\item Investigation of security needs that address the vulnerability of road transport introduced by the misuse of networked and co-operative systems. 
\item Integration of existing and emerging research and technology development (RTD) initiatives in order to support the introduction of security technologies in parallel to the progress of the technology infrastructure, and to ensure compatibility to legal and certification aspects.
\item Provision of qualified recommendations regarding (1) technology requirements (networks, architecture, systems and components and their interaction), (2) standardization needs, (3) legal provisions.
\end{itemize}

The WG will publish before the end of 2008 a report with two main parts. The first part will provide an overall view on security needs (motivation, prerequisites, state of the art, security analysis, use cases for security issues and security requirements). The second part will provide elements for
recommendations in the area of organization (e.g. related to the underlying public key infrastructure), of quality assurance and responsibilities aspects (e.g. related to the inspection needs). It  will also identify and prioritize research challenges. The report should lead to the implementation of concerted measures.

\section*{Car2Car Communication Consortium C2C-CC
Secure Vehicular Communication: Results and Challenges Ahead}

\noindent \emph{Benjamin Weyl, BMW Group Research and Technology}\\

Car2X communication enables a broad range of safety applications. Although this functionality inspires a new era of safety in transportation, new security requirements need to be considered in order to prevent attacks on these systems. Potential threats, security requirements and baseline security concepts of the C2C-CC Security Working Group are presented. One of the particular interests is trustworthy message exchange to ensure reliable, safe system operation, as well as the protection of identity and location against undesired privacy infringement. Different approaches are compared considering efficiency and scalability.

External communication interfaces, fixed and wireless, have increasingly become an integral part of automotive on-board architectures. This development is not the least driven by future safety application scenarios. Safety applications based on Car2X communication have been identified as a measure for decreasing the number of fatal traffic accidents. Examples for such safety applications are local danger warnings, traffic light pre-emption, or traffic information via road-side units. New security requirements need to be considered in order to prevent attacks on these systems. Attacks can be manifold: illegally forced malfunctioning of safety critical in-vehicular components and the illegal influence of traffic provoked by means of fake messages are just two likely possibilities.

\subsection*{Baseline}

Digital signatures are a convenient way to provide message integrity and authentication. Within the C2C-CC Security WG various approaches based on digital signatures have been discussed, outweighing the advantages and drawbacks with respect to the security and privacy requirements, as well as
scalability and performance constraints.

The use of long-lived pseudonym certificates fails to meet privacy requirements, because it makes vehicle and profile tracking possible, and node exclusion is not possible without the intervention of complex certificate revocation lists (CRL). The insufficiency of this approach could be fixed by using pseudonymous certificates pools, but it is desirable that a node does not own multiple simultaneously-valid pseudonyms in order to avoid Sybil attacks. Moreover, the revocation of such a pseudonym pool does not scale with a large number of nodes. Providing vehicles with a short-lived pseudonymous certificate, instead of several, solves the problem of the Sybil attack while keeping the benefits of the certificates pool approach~\cite{wey07}. Group signatures meet the privacy and scalability requirements, however, as computational effort is still too high, this mechanism is currently not applicable. Thus, the WG has chosen to advocate the use of short-lived pseudonymous certificates. Currently, the WG is discussing and specifying the appliance of this approach based on the C2C-CC reference model. More details on the baseline concepts and the activities of the WG can be found in the C2C-CC Manifesto~\cite{c2c-man}.

\subsection*{Challenges Ahead}

In order to prevent attacks where the in-vehicular system is tampered with (e.g., extracting secret material or manipulating the software), further security solutions are to be developed by combining software and hardware measures. More investigation is to be put into pseudonym change methods, change rates and certificate distribution. Besides technical aspects, other discussion areas are commercial requirements, regulation and legislation.

\section*{Securing Vehicular Communication Systems: Results and Next Steps}

\noindent \emph{Panos Papadimitratos, EPFL}\\

Our objective at EPFL, within the SeVeCom project~\cite{sevecom}, has been to design a baseline architecture that provides a sufficient level of protection for users and legislators and is practical and deployable. This baseline architecture is based on well-established and understood cryptographic primitives that can be implemented on today's hardware and deserve sufficient trust because of their existing deployment. It also allows deployed systems to be tuned or augmented, in order to meet more stringent future requirements.

The fundamental aspects that our \emph{architecture} seeks to address are: \emph{identity and cryptographic key management, privacy protection, and secure communication}. Additional problems to address are tamper-resistance and detection of faulty (inconsistent) data and node actions. In brief, we primarily seek to secure communications on the wireless part of the VC system, while protecting sensitive user information, and providing the option for node identification when necessary, e.g., for liability attribution. In other words, primary requirements are message authentication, integrity, and non-repudiation, as well as protection of private user information; a detailed discussion of the security requirements and the adversary models is available 
at \cite{kargl06,escar}.

Towards this end, we have developed an architecture~\cite{deliverable2.1,sevecom2007} that interoperates both the vehicular communication and the TCP/IP protocol stacks and relies on the presence of a Certification Authority (CA) and public key cryptography to protect V2V and V2I communication. Nodes, vehicles or road-side units (i.e., on-board vehicular communication platforms) are registered with CA, and each has a unique identity and is equipped with a pair of private and public keys and a certificate from the CA. These are long-term identities, credentials, and cryptographic keys. The $CA$ is also responsible for evicting nodes from the system, if necessary, either for administrative or technical reasons.

To enhance the user privacy, we rely on the concept of pseudonymity or \emph{pseudonymous authentication}: we require that each vehicle (node) is equipped with multiple certified public keys that do not reveal the node identity. The vehicle uses pseudonyms alternately, and each pseudonym for a short period of time without reusing it, so that messages signed under different pseudonyms cannot be linked. These short-lived keys are used to secure all communications, one- or multi-hop, with the senders and the relaying nodes of control or data packets signing and verifying them, depending on the employed protocol. Cryptographic and integrity protection prevent external adversaries from modifying and injecting traffic.

To protect the VC system from internal adversaries, i.e., misbehaving nodes equipped with the system credentials, we provide revocation methods. At first, as long-term credentials are used by vehicles to obtain new sets of pseudonyms, revocation can be performed at the entity providing the certification for new pseudonyms (which we can be in general different than the CA). Nonetheless, to revoke already granted and not expired credentials, we provide a Revocation of the Trusted Component (RTC) protocol, with the CA instructing the TC directly to erase all cryptographic material and acknowledge the termination of operation. In case RTC does not conclude successfully, we provide revocation through the distribution of compressed certificate revocation lists, namely, the RCCRL protocol, with RSUs acting as a gateway for the CRL dissemination and vehicles further relaying CRLs to other vehicles in parts of the network not covered by RSUs. To complement the eviction and enhance the system robustness against not-yet-revoked faulty nodes~\cite{jsac07}, we provide a localized misbehavior detection, a distributed self-protection and misbehavior evidence collection protocol.

However, the detection of misbehavior and its attribution to specific nodes is not an easy task in general. It may be feasible for specific types of deviation from the protocol (e.g., plausibility checks for geographic routing~\cite{har07}), but hard or even impossible for other types of misbehavior. Moreover, due to the highly volatile nature of vehicular networks, we cannot rely on lengthy interactions between two or more vehicles, in order to deduce the trustworthiness of specific vehicles. In fact, we realize this is a more general problem: We cannot operate exclusively on a priori or largely time-invariant trust relations with network entities. This is especially true if the identity of the data-producing entity is secondary, or if it is concealed by a privacy-enhancing mechanism. To address this challenge, we propose a shift towards \emph{data-centric trust}: Trustworthiness is attributed to node-reported data per se. We study this in the context of VC, with vehicles collecting reports (data), and we evaluate the trustworthiness of data reported by other vehicles rather than the trustworthiness of the vehicles
themselves~\cite{info08}.

Towards evaluating and enhancing the practicality of our secure communication architecture and protocols, we consider the simplification of the management of the short-lived cryptographic keys
(pseudonyms) and credentials and the satisfaction of privacy and security requirements. To achieve this, we propose a \emph{Hybrid Scheme} \cite{Calandriello2007}: This is essentially a pseudonymous authentication scheme that enables nodes to generate their own certified pseudonyms. To maintain the degree of privacy protection pseudonymous authentication provides, nodes utilize a group signature (GS) scheme. Thus, the most important aspect that the Hybrid Scheme changes with respect to the ``baseline'' pseudonymous authentication is the computation of pseudonyms and certificates and, 
consequently, their validation. Such a scheme is modular, usable, efficient, and robust. It eliminates the need for pre-loading, storing and refilling pseudonyms and the corresponding credentials and private keys, so that vehicles do not need to be either side-lined, or they are not forced to compromise their user's privacy if insufficient or no pseudonyms are available, or they do not need to ``over-provision'' their pseudonym supply.

We also investigate how to reduce the cost due to security and privacy enhancing technologies. In particular, we look at variants of pseudonymous authentication, as discussed above, and we introduce a set of optimizations to reduce processing and communication overhead. But we are interested in a more general problem: What is the effect of security and these broadly accepted pseudonym-based mechanisms on transportation safety? In other words, can, for example, vehicle collisions still be avoided, thanks to VC-enabled safety applications, even if security is integrated? We set out 
to answer this question by evaluating the reliability of communication, the processing load at each node and the overall impact on the transportation safety, expressed as the proportion of collided vehicles, under various conditions, by binding traffic and network simulation. Our results show the need for increased processing power, as well as the benefits from the proposed optimizations and
the ability to achieve safety practically at the same level as that achieved by an unsecured emergency braking alert application~\cite{move08}.

With the knowledge about how to address a number of fundamental issues, and the appropriate tools to evaluate our schemes, we have several encouraging results and approaches for designing and deploying practical secure and privacy-enhancing VC systems that also rely on non-cryptographic defense mechanisms. These systems could then achieve essentially the same level of transportation efficiency and safety as a system that would operate in a benign environment without security.

\section*{PriPAYD: Privacy Friendly Pay-As-You-Drive Insurance}

\noindent \emph{Bart Preneel, K.U. Leuven}\\

Pay-As-You-Drive (PAYD)~\cite{TDKP-PriPAYD} is a new car insurance model where, in contrast to the traditional pay-by-the-year policy, customers are charged depending on where and when they drive instead of a fixed amount per year. For each kilometer that a car is driven the statistical risk of accident, depending on the road and the time of the day, is calculated and translated to
personalized insurance fees. Pay-As-You-Drive insurance models are hailed as the future of car insurance due to their advantages for users and companies~\cite{Litman,Zahid}. First, the insurance fees applied to each user are fairer than those in the pay-by-the-year scheme, as customers are only charged for the actual kilometers they travel. Second, PAYD policies are socially and environmentally beneficial, as they encourage responsible driving, decrease the risk of accidents (which in turn saves money for users and insurers) and reduce energy consumption and pollution emissions. Due to all these advantages, PAYD insurance policies are being widely developed by 
insurance companies all over the world like Norwich Union~\cite{NU} (UK), Aioi~\cite{Aioi} (Japan), Hollard Insurance~\cite{Hollard} (South Africa), etc. 

Although PAYD insurance seems to have many advantages, most of its current implementations involve an inherent threat to user's privacy. The full information used for billing (the time and location of the car) is gathered by a black box in the car, and transferred to the insurance company (and, in some of the cases, to a third company providing the location infrastructure). The insurer does the accounting to obtain the client's premium and sends the bill by traditional mail, together with a user friendly (reduced) version of the full GPS data. This model puts service providers in a business advantage position. With all the data collected, new services (traffic information, 
pollution information, etc) can be offered to customers. It also allows providers to perform data mining to detect potential fraud. However, the obvious disadvantage of this model is that it is privacy invasive, as the data collected by the insurance company is sufficient to track almost every movement of a car over time. Moreover, this collection of personal data raise many legal questions, as pointed out in~\cite{TDKP-PriPAYD}.

We propose \PPAYD\, a privacy friendly scheme. Our proposed architecture follows closely the 'current model' with the exception that the raw and detailed GPS data is never provided to third parties. Computations transforming the GPS data into billing data are performed in the vehicle black box and the insurer receives only the billing data instead of the exact vehicle locations (thus they cannot invade the user's privacy) while being sure he is receiving the correct data. The client can check that only the allowed data is in the insurance company database, and the raw data is available for the client (or a judge) to check the correctness of the bill. Our techniques also permit easy
management and the enforcement of the policies by the insurer. 

Periodically, the premium for a period of time is calculated and the amount to be payed is sent in a secure way to the insurance company via GPRS, or even the less expensive SMS services. The data is signed using the black box key and encrypted under the public key of the insurance company. To ensure that the black box is not acting maliciously in favor of the insurance company, we need to allow a car user or owner to audit the billing mechanism. For this purpose, we propose the use of an off-the-shelf USB memory stick. The data is recorded in an encrypted way on this token so that only the policy holder can access it, and it is signed by the black box to be usable as evidence. The symmetric encryption key is generated by the black box and provided to the policy holder in two key shares: one written on the USB stick and the other relayed through the insurance company and delivered by mail with the bill. A simple mechanism, such as pushing a button on the box for some time, allows the encryption key to be reset.

There is no component or infrastructure required by \PPAYD\ that would make it much more expensive than current systems. One could in fact argue that in the long run running \PPAYD\, as any other privacy enhanced technology, is less expensive than privacy invasive systems. The costs of protecting private data stores is often overlooked in the accounting of costs, as is the risk of a single security breach leaking the location data of millions of policy holders. In addition, \PPAYD\ keeps sensitive data locally in each car, in a simple-to-engineer-and-verify system. Requiring an off-the-shelf back-end system to provide the same level of privacy protection to masses of data would make them, not only prohibitively expensive, but simply not implementable.

\section*{Panel: ``Secure vehicular communications: What are the main research challenges left?''}

\noindent \emph{Chair: Jean-Pierre Hubaux \\
Panelists: Emilio Davila-Gonzalez (EU Commission), Albert Held
(Daimler), William Whyte (NTRU) \\
Report: Maxim Raya and Jean-Pierre Hubaux}\\

According to William Whyte, misbehavior detection and revocation in vehicular communications remain open problems. The certification of vehicular applications (i.e., the end points of communication in the IEEE 1609 draft standard) is another unsolved problem. Most interesting is the panelist's statement that US car-makers favor anonymity of drivers over their liability in case of accidents. Hence, misbehavior would be reported to the CA (Certificate Authority) only in critical cases and revocation would be carried out only after a long history of misbehavior. The reason for favoring anonymity over driver liability is the car-makers' fears of being sued if data obtained from
vehicular communications causes legal or financial damage to drivers (a typical example being that of an affair revealed by a private detective, based on the vehicular communication traces of the suspected spouse's vehicle).

Albert Held summarized the remaining challenges in three components: prevention, detection, and recovery. The main issue in prevention is the possible lack of a PKI (Public Key Infrastructure) and of a CA, at least in the early stages of deployment. The typical services provided by the PKI, such as key management, need to be substituted by intermediate solutions until the proper infrastructure is available. In detection, the hard question is: What is an attack? It is often hard to distinguish between a malicious error and a faulty device. Hence, a related question is whether a faulty device should also be considered to be an attack because it causes damage to the system. Last but not least, recovery from attacks is also tightly related to the notion of fault tolerance. Based on the attack or fault, the recovery process should either remove the attacker or repair the fault.

Emilio Davila-Gonzalez described the ongoing and future EU proposals related to transport safety and security. In his opinion, the remaining challenges are the integration of the security architecture into a harmonized EU-wide vehicular communication architecture and the implementation of the proposed security solutions in the planned operational field tests of cooperative vehicular
systems.

In addition to the panelists, the audience highlighted the following set of additional challenges:
\begin{itemize}
\item Specification of
proper requirements for the security of vehicular communications.
\item Information aggregation primitives for reducing the overhead of
information dissemination, considering that security might add a high
overhead to communication.
\item  Making security cooperative, among
vehicles, and reactive instead of the current local, to the vehicle,
and passive approach (e.g., like anti-virus software).
\item Secure positioning, given that both GPS and Galileo lack the necessary
security mechanisms.
\item To the question about the existence of
real security threats to vehicular communications, the panelists unanimously
pointed out that the main motivation of car makers are terrorist attacks,
especially in the US, and the kidnapping threats for luxury brands.
\end{itemize}

An audio recording of the panel, courtesy of Thomas Heydt-Benjamin, is available at: http://www.archive.org/details/\\CryptocracySpecialEpisode001EpflWorkshopPanel.

\section*{Security for Inter-Vehicular Communication Mechanisms: What is next?}

\noindent \emph{Frank Kargl, Elmar Schoch, and Michael Weber, Ulm
University}\\

In vehicular communication systems considered by many research projects (e.g., Fleetnet, VSC, Network-on-Wheels, VII, CVIS, Safespot) and standardization groups (e.g., IEEE 802.11p and 1609.x, ISO-CALM, Car-2-Car Communication Consortium) we have identified a recent trend away from classic
communication patterns and towards more sophisticated forms of communication.

Earlier research mostly addressed the following communication patterns:

Beaconing: direct and periodic broadcast of messages to all neighbors reachable
via the wireless radio.

Flooding and Geocast: (Potentially also periodic) distribution of broadcast
messages where receivers act as relays. Distribution is usually restricted by
time-to-live (TTL) counters or a geographic destination area in case of Geocast

Position-based Routing: In contrast to topology-based routing that is often used in MANETs, position-based routing has proven to be superior for vehicular networks.

Up to now, these mechanisms are mainly regarded in work on security and privacy of vehicular networks. Recently, research on vehicular communication has begun to suggest more advanced means of information dissemination and this, of course, necessitates an adaption of security and privacy mechanisms. Before addressing those issues, we first give some examples of such communication mechanisms.

Various publications highlight the need for more efficient flooding and Geocast strategies. Depending on network parameters like node density or topology, e.g., Gossiping decreases the probability by which a node relays a received packet. This can lead to significant increase of network efficiency.

The so-called Context-adaptive Message Dissemination introduces the idea of contextual relevance for the dissemination of information. Based on parameters such as the source location or the age of an information, the node estimates the relevance of an information to its neighboring nodes and preferably forwards messages with higher relevance. A modified medium access scheme even allows
inter-node prioritization based on this relevance. Overall, the available bandwidth is used primarily to forward information that is important in the current context, whereas less relevant information is delayed or discarded. 

Aggregation goes even one step further. When a node receives data from neighboring nodes, it does not immediately forwarded them. Instead a node checks if it can aggregate this information with other information received earlier or generated locally. E.g., in a traffic jam, many nodes will report similar data, e.g. low speeds. This can easily be aggregated, to directly reduce the amount of communicated data.

From a security and privacy point of view, Gossiping, Context-adaptive Message Dissemination, and Aggregation, by themselves already, provide an astonishing degree of resistance against attacks. In contrast to many routing protocols, there is (almost) no signalling between nodes that an attacker could exploit. Essentially an attacker is limited to Denial-of-Service attacks or modification/forging of information.

This however can hardly be addressed by traditional, crypto-based security mechanisms alone. Those mechanisms often provide a sender-centric security where the sender of a message protects it from modification by means of signatures or from eavesdropping by encryption. Furthermore, those mechanisms often assume a mostly static packet content that is disseminated in the network with only minimal changes, e.g. decreasing a TTL in the packet header.

Whereas this assumption might hold in the case of Gossiping, information may already be re-arranged to new packets in the case of Context-adaptive Message Dissemination. When using aggregation mechanisms, individual information usually is lost during the dissemination process.

Hence, the sender- and packet-oriented approach to security needs to be replaced or augmented by a data-oriented approach where mechanisms like consistency-checks using redundant information or real-world sensors are used to discard incorrect information from the network and where rate-limits
restrict the effects of Denial-of-Service attacks. Initial exploration of such mechanisms has already delivered promising results.

\section*{Secure and Privacy-Enhanced Geocast: Results and Challenges}

\noindent \emph{Andreas Festag, NEC Germany GmbH, Heidelberg, Germany} \\

Geocast is a network protocol for ad hoc and multihop communication with short-range wireless technology, such as IEEE 802.11. It utilizes geographical positions for addressing and packet routing and provides various forwarding schemes, including geographical unicast, broadcast and anycast. It quickly adapts to frequent network topology changes and naturally supports the
distribution of data packets in geographical target areas. Hence Geocast is particularly suited for vehicular networks targeting at road safety and
travel comfort applications. For a sustainable deployment of Geocast in
realistic environments, security and privacy are inevitable components.

Security objectives for Geocast cover integrity, authentication, and non-repudiation of packet header and payload, including: (i) cryptographic protection based on digital signatures and certificates, (ii) plausibility checks of fields carried in the network header and their local confidence assessment, (iii) trustworthy forwarding and network isolation compromised nodes, and (iv) rate control to prevent forwarding of massively injected data. For the cryptographic primitives, we distinguish between immutable and mutable header fields, which indicate whether the fields can be changed in the forwarding process or not, and combine hop-by-hop and end-to-end signatures to protect the mutable and immutable fields, respectively.

Privacy ensures that a node - source, forwarder, and receiver - is not identifiable. It hides personal data such as location, speed, and heading, but it allows a node to reveal its identity to other nodes for reasons such as reputation and session establishment or for legal authorities. The core concept for achieving privacy is based on the use of pseudonyms. In order to prevent tracking, a node changes pseudonyms frequently. Privacy-enhanced Geocast implies solutions for (i) a cross-layer addressing concept for MAC, GeoCast and IPv6 addresses derived from pseudonyms, (ii) control of effective pseudonym changes, (iii) mitigation of the impact by pseudonym changes on routing, and (iv) a pseudonym resolution service for (re-)establishment of communication sessions.

\begin{figure}[ht]
\centering
\includegraphics[width=\linewidth]{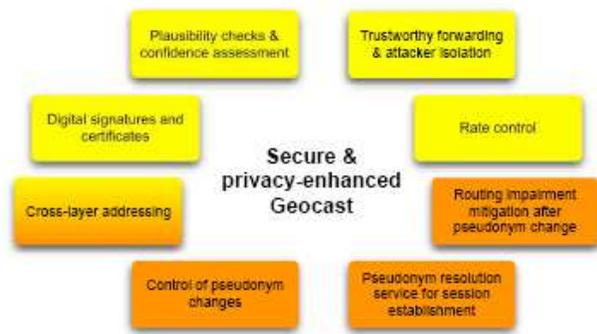}
\caption{Main components} \label{fig:af}
\end{figure}

The solution for secure and privacy-enhanced Geocast is implemented in a software prototype for vehicular communication based on IEEE 802.11~\cite{nec-sdk} and serves as the basis for ongoing and future R\&D efforts and projects. As complements to the existing solution, a number of challenges can be identified, from which we highlight these:
\begin{itemize}
\item \textbf{Hardware acceleration} for cryptographic operations in order to meet real-time delay
requirements of road safety applications~\cite{har07} and to allow for an advanced forwarding scheme that presume a negligibly small processing delay (e.g., contention-based forwarding).
\item Integration of communication protocols for vehicular networks with \textbf{wireless
sensor networks}\footnote{We regard WSN as a cost-efficient infrastructure for traffic safety, which lowers barriers for ITS deployment.} (WSN), secure the aggregation and distribution of data, and secure access to stored information in the sensors.
\item Deployment of advanced PKI concepts with inherent support of user privacy, as described in~\cite{arm07}.
\end{itemize}

\bibliographystyle{plain}
\bibliography{bib-mc2r}

\end{document}